\documentclass[11pt]{article}

\usepackage{SIGACTNews}

\newcommand{\COLUMNsplit}{\vskip 2em\hrule \vskip 3em}

\newcommand{\COLUMNguest}[2]{%
  \begin{center}%
   {\LARGE\bf #1 \par}%
    \vskip 1.5em%
    {\Large
      \lineskip .75em%
      \begin{tabular}[t]{c}%
        #2
      \end{tabular}\par}%
    \vskip 1.5em%
  \end{center}%
  \par}

\usepackage{amssymb,proof,code}

\title{SIGACT News Logic Column 16}
\author{Riccardo Pucella\\
Northeastern University\\
Boston, MA 02115 USA\\
riccardo@ccs.neu.edu}
\date{}

\begin{document}

\SIGACTmaketitle

In this issue, Karl Crary and Robert Harper respond to the critique of
higher-order abstract syntax appearing in Logic Column 14, ``Nominal
Logic and Abstract Syntax'' (SIGACT News 36(4), December 2005).

I am always looking for contributions. If you have any suggestion
concerning the content of the Logic Column, or if you would like to
contribute by writing a column yourself, feel free to get in touch
with me. 

\COLUMNsplit

\COLUMNguest{Higher-Order Abstract Syntax: Setting the Record Straight\footnote{\copyright{} Karl Crary and Robert Harper, 2006.}}
   {Karl Crary \qquad Robert Harper\\ Carnegie Mellon University}

A recent SIGACT News Logic column, guest-written by James Cheney,
discussed {\em nominal logic\/}~\cite{gabbay+:nominal-logic}, an
approach to abstract syntax with binding structure.  In addition to
providing a worthy tutorial on nominal logic, that column leveled five
criticisms at {\em higher-order abstract syntax,} an alternative
approach for dealing with binding structure in abstract syntax.  We
argue below that three of those criticisms are factually inaccurate,
and the other two are misguided.

Since higher-order abstract syntax (HOAS) is a general concept, not a
specific formalism, it is difficult to make precise statements about
it.  Therefore, we will consider the column's claims in the context of
the Logical Framework (LF)~\cite{harper+:lf}, the best-known and
best-developed realization of higher-order abstract syntax.

\begin{enumerate}
\item
The column claims that HOAS is based on complicated semantic and
algorithmic foundations, and mentions explicitly higher-order logic,
recursive domain equations, and higher-order unification.  None of the
three are necessary to understand LF:

\begin{itemize}
\item
All the metatheory of LF can be (and typically is) carried out in
first-order logic; higher-order logic is not necessary.

\item
The metatheory of LF is carried out using entirely syntactic means.
There is no need to resort to any denotational semantics, including
recursive domain equations.

\item
While higher-order unification is used in the
Twelf implementation~\cite{pfenning+:twelf-manual} of LF to support
type inference and proof search, it is not at all relevant to the
methodology of encoding abstract syntax in LF.  Moreover, even in
regards to the Twelf implementation, experience suggests that
understanding the higher-order unification algorithm is not essential;
students who do are not familiar with it nevertheless have little
difficulty in using Twelf.
\end{itemize}

\item
The column claims that structural properties of the meta-language must
be inherited by the object language.  This is simply untrue.  For
example, linear logic can be adequately represented in ordinary LF
using a linearity judgement; one need not resort to a linear variant
of LF.

\item
The column states that variable names in LF ``have no existence or
meaning outside their scope.''  This is true, in the sense that
variable names alpha-vary, as is standard practice in logic and
programming languages.  The fact that alpha-equivalence is built-in is
a strength of HOAS (and of de Bruijn representation) not enjoyed by
first-order abstract syntax or by nominal logic.  One might wish to
argue that the usual practice of considering terms modulo
alpha-equivalence is counter-productive, but if so, the objection is
with the field in general, not with HOAS.

Moreover, taking alpha-equivalence as built-in does not preclude the
isolation of variables in LF (which can be done easily using a
hypothetical judgement).  Consequently, it is not difficult to encode
logics that deal with variables specially or that compare free
variables for equality.

The real contrast between HOAS and nominal logic lies in their
philosophy toward binding structure.  Both HOAS and nominal logic
provide a sophisticated treatment of binding structure not enjoyed by
first-order accounts, but ones of very different character.  The
philosophy of LF is that binding and alpha-conversion are fundamental
concepts, and thus builds them in at the lowest level.  Consequently,
name management in LF is ordinarily a non-issue.  In contrast, nominal
logic promotes name management as an object of study.  It makes
binding into a sophisticated artifact, rather than a primitive concept
as in LF.

\item
A fundamental part of the LF methodology is the notion of an adequate
encoding.  Syntactic terms and derivations of judgements are
represented in LF as certain canonical forms, and the encoding is said
to be adequate if there exists an isomorphism (in a sense that can be
made precise) between the terms/derivations on the object language and
the canonical forms that represent them.  This provides a clear
criterion for whether an encoding is correct or not.

The column notes that inadequate encodings have appeared in papers,
and takes that fact as evidence that it is too hard to develop
adequate encodings.  To the contrary, we see the existence of
incorrect encodings as merely the inevitable consequence of the
existence of a correctness criterion.

To the best of our knowledge, nominal logic does not currently enjoy a
precise criterion for correctness.  Consequently, it is hard formally
to justify the statement that anything at all has been encoded in
nominal logic.  When nominal logic is better understood, we expect
that a correctness criterion will be developed for it as well.  Once
that happens, we expect that that criterion too will be used to
identify incorrect encodings.

\item
The column claims that HOAS cannot address ML-style let-polymorphism
due to difficulties with open terms.  This is not so.  There are at
least two perfectly satisfactory formulations of let-polymorphism in
LF.  The simplest is to re-typecheck the let-bound variable at each
reference:

\begin{code}
of-let        : of (let E1 ([x] E2 x)) T
                 <- of E1 T'
                 <- of (E2 E1) T.
\end{code}

If one prefers a formulation with an explicit account of generalization,
that is also possible:

\begin{bigcode}
 polytp : type.
 mono   : tp -> polytp.
 forall : (tp -> polytp) -> polytp.
 \mbox{}
 of     : term -> tp -> type.      %
 polyof : term -> polytp -> type.  %
 assm   : term -> polytp -> type.  %
 \mbox{}
 polyof-mono   : polyof E (mono T)
                  <- of E T.
 \mbox{}
 polyof-forall : polyof E (forall ([t] S t))
                  <- ({t:tp} polyof E (S t)).
 \mbox{}
 of-let        : of (let E1 ([x] E2 x)) T
                  <- polyof E1 S
                  <- ({x:term} assm x S -> of (E2 x) T).
 \mbox{}
 ...
\end{bigcode}

In the alternative formulation alluded to by the column, the let-bound
term is given a type (possibly including free variables) which is then
forcibly generalized by a polymorphic closure operator.  The inherent
clunkiness of such a formulation does makes it awkward to encode in
LF.  However, the difficulty does not result from its use of open
terms, which are easily expressible:

\begin{code}
openterm : type.
trm      : term -> openterm.
close    : (term -> openterm) -> openterm.
\end{code}

\end{enumerate}

LF does lends itself particularly to languages enjoying hygienic
scoping disciplines.  Although nothing prevents one from employing
first-order representations in LF, to do so can sacrifice much of the
advantage afforded by LF\@.  Consequently, nominal logic looks very
attractive for languages with more novel notions of scope.  However,
for languages with largely conventional scoping, LF provides an
extraordinary tool that is being used by a growing number of
researchers on a daily basis to formalize their metatheoretic results.

\bibliographystyle{plain}

\end{document}